\documentstyle[jkas]{article}
\runningauthor{HESSER ET AL.}
\runningtitle{FORMATION OF THE MILKY WAY }
\beginpage{1}
\endpage{8}

\begin{document}

\title{FORMATION OF THE MILKY WAY}

\author{J. E. ~Hesser,$^1$
P. B. ~Stetson,$^1$
W. E. ~Harris,$^2$ 
M. ~Bolte,$^3$
T. A. ~Smecker-hane,$^6$
D. A. ~Vandenberg,$^4$
R. A. ~Bell,$^5$
H. E. ~Bond,$^7$
S. ~Van~den~bergh,$^1$
R. D. ~Mcclure,$^1$
G. G. ~Fahlman,$^8$ and
H. B. ~Richer$^8$}

\address{$^1$National Research Council of Canada,
 Herzberg Institute of Astrophysics,
 Dominion Astrophysical Observatory} 
\address{$^2$McMaster University} 
\address{$^3$Lick Observatory}
\address{$^4$University of Victoria}
\address{$^5$University of Maryland}
\address{$^6$University of California, Irvine}
\address{$^7$Space Telescope Science Institute}
\address{$^8$University of British Columbia}

\abstract{ We review observational evidence bearing on the formation of a
prototypical large spiral galaxy, the Milky Way. New ground- and
space-based studies of globular star clusters and dwarf spheroidal
galaxies provide a wealth of information to constrain theories of galaxy
formation. It appears likely that the Milky Way formed by an combination
of rapid, dissipative collapse and mergers, but the relative contributions
of these two mechanisms remain controversial. New evidence, however,
indicates that initial star and star cluster formation occurred
simultaneously over a volume that presently extends to twice the distance
of the Magellanic Clouds.}

\keywords{astronomy, astrophysics}

\maketitle  

\section{ OVERVIEW }           

It is an honour to open this meeting,  with its enormous 
range in scope from the solar system to the furthest realms by
addressing a topic in which our Korean hosts are making so many
important contributions.  The goals of this review are to examine,
however fleetingly, six themes, and thereby to impart a flavor of
issues and challenges currently debated by those intrigued by how 
the Galaxy, and especially its halo, formed some 15~Gyrs ago.  
The themes are encapsulated within six whimsical questions:

\begin{enumerate}
\item Why should we care?
\item Are we in Jurassic Park?
\item Can you trust anyone over 10~Gyrs old?
\item Star formation by magic in the outer halo?
\item Have we got the right spices in the kimch'i?
\item Do you see that ghost?
\end{enumerate}

\section {QUESTION 1}

Justifiably, much of current astrophysics aims to understand the
distribution of matter in the Universe and the formation of galaxies by
studying high redshift objects. Many remarkable intersections of
enlightenment may, however, be found between detailed studies of the Galaxy
and research on even the largest scales or highest redshifts in the
Universe. In turn, these intersections -- some of which we will
mention in this section -- make the question of the formation
of the Milky Way a vitally interesting subject in present-day
astrophysics.

In a critical review, Fukugita, Hogan and Peebles (1996) assemble
observational constraints on galaxy formation theory. In a cartoon (their
Figure 5), they argue that spheroids of large galaxies were likely formed
by redshift z = 3, and that disks were assembled at redshifts 1$<$z$<3,$
by which time most gas had turned into stars.  It is interesting to
consider how much this view contrasts with currently popular hierarchical
merger models based upon N-body simulations (e.g., Lacey and Cole 1993,
Cole and Lacey 1996), in which the halo of a large galaxy like ours is
built by mergers of many sub-galactic units. Trying to
understand the relative importance of mergers and dissipative collapse is
a goal of many active programs on Galactic research. The debate usually
compares strengths and weaknesses of Eggen, Lyden-Bell and Sandage's
(1962; see also Sandage 1990) vision for Galactic formation via a rapid
collapse with accompanying star and star cluster formation, and Searle and
Zinn's (1978) mergers over many Gyr of smaller galaxies that have
undergone independent evolution before blending into the Milky Way.

While most attempts to model galaxies assume spherical geometry for the
distribution of the enigmatic dark matter, a detailed analysis (Hartwick
1994, 1996) of the spatial and kinematical properties of stellar systems
in the Galactic halo suggests that they define a cigar-shaped potential.
If true, how will this affect the formation and evolution of individual
galaxies and of groups like the Local Group?

Today much of astronomy wrestles with the conflicting ages implied by
recent measures of the cosmic distance scale and those from a
comparison of globular star cluster color-magnitude diagrams (CMDs) to
stellar evolution theory. Do we have a crisis in cosmology? in stellar
evolution theory? in distance indicators? in all? or are the
current discrepancies mere artifacts of overly optimistic error bars?

Finally, is it not fascinating that in the Galactic halo (z = 0) we find
stellar systems whose chemical abundance ratios overlap with high
redshift Lyman-$\alpha$ clouds (e.g., Pettini, et al. 1994), thus
permitting detailed study of stars from an epoch long vanished? For
these and many other reasons, study of the formation of the Milky Way
remains an endeavor vitally relevant to the central physical questions
of our era.

\section {QUESTION 2}

Much of this review deals with globular clusters in the Milky Way
halo, for which there is evidence (e.g., Richer, et al. 1996) that,
with a few exceptions, they may be uniformly very old with little or no
signs of an age-metallicity relation (this is a controversial
subject, as the references in Richer et al. detail).  But, are
globular cluster systems always relics of the earliest epoch of galaxy
formation, or are at least some systems actively under construction
today? For instance, do multiple peaks in the color distributions
(usually equated as representing peaks in the metallicity distribution
function) of cluster systems around massive galaxies (e.g., Geisler,
Lee \& Kim 1966 for the giant elliptical, M\thinspace49) represent
multiple epochs of cluster formation that, in turn, argue that most
large galaxies were predominantly formed by accretion or merger
processes, perhaps extending over a major fraction of their
lifetime?

Another constraint on the relative epochs of star formation in large
galaxies arises from the fact that globular clusters in E and dE galaxies
appear to be more metal-poor than the underlying field halo stars, which
suggests that most of the globular clusters formed before the halo stars
(Harris 1996).

Evidence continues to build (e.g., Hilker \& Kissler-Patig 1996, Holtzman,
et al.\ 1996, and references therein) that super-starclusters forming now
in galaxies undergoing major bursts of star formation (e.g.,
NGC\thinspace3597, NGC\thinspace6052, and NGC\thinspace7252) have
luminosities and colors consistent with their being young, massive
globular clusters.  Even more convincing evidence is beginning to come
from dynamical mass measurements: the velocity dispersions that Ho \&
Filippenko (1996) derive for super-starclusters in NGC\thinspace1569 and
NGC\thinspace1705 are as large as expected if the clusters are analogs to
globular clusters.  Curiously, few such objects are found in cooling flow
systems.  Harris, Pritchet \& McClure (1995) found no correlations between
properties of the cluster systems they studied and those of cooling flows.
They argue that recent cluster formation most likely results from sporadic
starburst activity, and that larger cluster systems in cD galaxy halos
stem from an intensive phase of cluster formation in the protogalactic
epoch.  Thus, some objects forming now (in heavenly
`Jurassic Parks') should fade in roughly a Hubble time to become analogs of
present-day Milky Way globular clusters. It is far from clear, however,
that a significant fraction of any galaxy's star cluster system comes from
any but an ancient era, nor whether systems with some recent cluster
formation will evolve to resemble the globular cluster system of the Milky
Way today.

Taken together, the recent discoveries of young analogs for the ancient
globular star clusters familiar to us all from the Milky Way's halo
suggest that present-day cluster formation efficiency is low. Thus,
recent cluster formation is not likely a panacea for building the
populous globular cluster systems belonging to large ellipticals.
Consequently, van den Bergh's (1990) long-standing concern remains,
namely that it seems extremely difficult to build the populous clusters
systems of large ellipticals predominantly by mergers of spirals, with
their more modest cluster systems.

\section {QUESTION 3}

In this and the next section we turn to age determinations for stellar
systems in the Galactic halo, which is the focus of much of our current
research. Several comprehensive reviews (e.g., Bolte and Hogan 1995;
VandenBerg, Stetson and Bolte 1996; Stetson, VandenBerg and Bolte 1996)
are drawn upon here. There are two broad issues to address:  a) how old
(in absolute terms) are the oldest globular clusters, the answer to which
sets a lower limit to the age of the Universe; and b) is there a range of
ages among the Galactic globular clusters and, if so, what does it
say about how the Milky Way formed?

Bolte and Hogan's (1995) Figure 2 summarizes well the current situation
with regard to absolute ages for the most metal-deficient star clusters in
the Galactic halo. Using the models of Bergbusch and VandenBerg (1992) for
physical parameters thought to represent best the classical halo cluster
M92 ([Fe/H]$=-2.26$, Y = 0.235, [O/Fe]$=+0.7$), and employing critically
assembled values for solar-neighborhood subdwarfs, they conclude that M92
has an age of 15.8$\pm2.1$~Gyrs.  Were helium diffusion included in the
models, the inferred age would be $\sim$15~Gyr. If the Hubble parameter,
H$_{\rm o}$, is of order 70 km/sec/Mpc in a standard inflationary
cosmological model with a zero cosmologial constant and $\Omega<0.2$ (see,
e.g., Carlberg, et al.  1997), then there is a $\sim$3$\sigma$ discrepancy
between ages inferred from globular clusters and from H$_{\rm o}$ (see,
for example, Figure 3 of Freedman, et al. 1994):  hence, the perceived
crisis in observational cosmology.

We are excited by the prospect that it now appears possible with the
Hubble Space Telescope (HST) to use, for the first time, a quite
different stellar chronometer to estimate globular cluster ages, and
thus to provide an independent check on the traditional turnoff
luminosity technique. This method compares the white dwarf luminosity
functions in the nearest globular clusters, such as M4 (see Richer,
et al. 1995, 1997), to theoretical white dwarf cooling curves. HST
data in hand suggest that, if the M4 white dwarfs have carbon cores,
then the faintest stars observed have ages of $\sim$8~Gyrs, which would
be incompatible with standard inflationary cosmological models for
H$_{\rm o}>83$. Moreover, the inferred ages of the faintest M4 white
dwarfs are already within 2~Gyrs of the oldest known disc white dwarfs
in the solar neighborhood, 10.5$\pm\sim2$~Gyrs (Oswalt, et al. 1996).

Relative ages for objects in the Galactic halo are rather easier to
measure than absolute ages; nonetheless, relative ages are powerful
probes of how the Galaxy might have formed. Two techniques for
determining relative ages, each with strengths and weaknesses (Salaris,
Chieffi \& Straniero 1993), are regularly applied in the literature.
(1) The luminosity difference between the horizontal branch (HB) and
main-sequence turnoff (TO), $\Delta$V, and (2) the color difference,
$\Delta$(color) (typically, B$-$V or V$-$I) between the TO and the base
of the giant branch. Theory predicts luminosities more reliably than
colors, but method (1) is operationally more difficult to apply than
method (2), and suffers from uncertainties about the dependence of HB
absolute magnitude on metallicity, among other things.  Method (2), on
the other hand, cannot be easily applied to objects with [Fe/H]$>-1.0$,
as described by VandenBerg, Bolte \& Stetson (1990) or Sarajedini \&
Demarque (1990).

\begin{figure}[t]
\centerline{\epsfysize=10cm\epsfbox{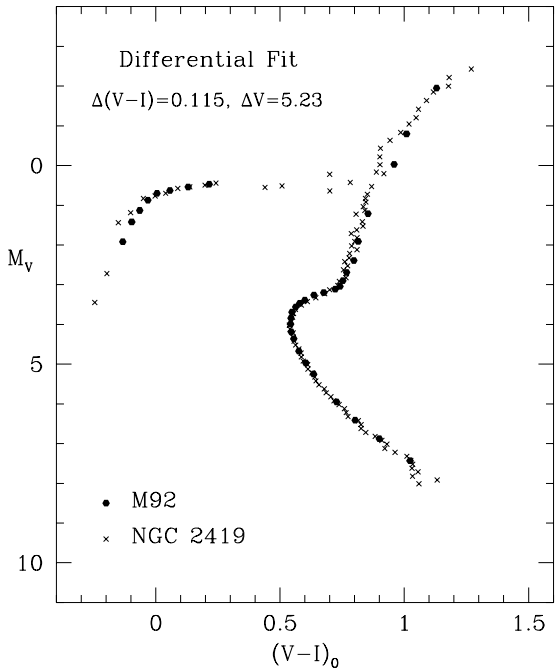}}
\caption{Comparison of the principal sequences of M92 and NGC\thinspace 2419.}
\end{figure}

A third method, developed by Lee, Demarque \& Zinn (1994; hereinafter,
LDZ), uses synthetic HB models to analyze the distribution of stars on the
HBs of globular clusters. The HB phase of stellar evolution is extremely
sensitive to changes so small as to be below the limits of present
observational technique of essentially every parameter of which one can
think (e.g., $\Delta$Y, $\Delta$[C,N,O/Fe], mass loss, core to envelope
mass, etc.).  Through a process of elimination, LDZ provide a scenario
(see their Figure 7) in which differences in age between clusters in
different regions of the halo provide a consistent picture. For example,
for Galactocentric distances R$<8$~kpc, LDZ's model suggests ages that are
about 2~Gyrs older than for clusters with 8$<$R$<40$~kpc, which in turn
are about 2~Gyrs older than clusters with R$>40$~kpc. Moreover, the inner
halo appears in their picture to be more uniform in age than does the
outer halo, where the puzzling `second-parameter' effect (redder HBs than
expected for the [Fe/H]) becomes more evident. [Note also that the
second-parameter effect has been observed among globulars of the Fornax
dSph (Smith, et al.  1996), so it is not peculiar to the Milky Way.] LDZ
thus conclude from their analyses that the halo of the Galaxy was
assembled in the manner envisaged by Searle \& Zinn (1978) over something
like one third the life of the Galaxy, with the second parameter
identified predominantly with differences in age.  Using method 1 on a
hetrogenous data set, Chaboyer, Demarque \& Sarajedini (1996) also infer
age to be the second parameter. This picture's simple elegance has led to
its becoming widely accepted. While no doubt age differences between halo
clusters is part of the answer, recent careful differential comparisons
suggest that Nature may not yet have fully revealed her secrets to us.

For instance, Stetson (1995) reported highly accurate differential
photometry (same night, telescope, detector, analysis, etc.) for the
canonical Northern hemisphere second-parameter pair, M3 and M13. From
the CMDs in the TO and subgiant regions, he finds that M13 can be at
most 0.5~Gyr older than M3 and/or M13 is 0.5~dex more metal poor than
M3. However, from high-dispersion spectroscopy, Kraft et al. (1993)
find that the two clusters have the same [Fe/H] to within
$\pm$0.04~dex, while the synthetic HB studies implied they differ in
ages by more than 2.5~Gyr.  Because there is evidence for deep mixing
in M13 in the form of six first-ascent giants that are very deficient
in oxygen, Kraft et al. speculate that perhaps stellar angular momentum
is regulating the mixing and driving the HB morphology, rather than
differences in ages.

We have been using the Hubble Space Telescope (HST) in Cycles 4-6 to
make V,I CMDs for objects in the outer halo of the Galaxy. Our goals
include absolute and relative age determinations that we hope will
illuminate the formation scenario for the Milky Way. Most of the
clusters in the outer halo have red HBs and are of low luminosity.  An
exception is NGC$\thinspace$2419, a luminous, blue HB cluster located
at R$_{\rm gc}$=95~Kpc (i.e., nearly twice the distance of the Large
Magellanic Cloud).  NGC$\thinspace$2419 has the same
[Fe/H] as the inner halo cluster, M92. From M$_{\rm v}=+8$ to $-2$ the
two CMDs are indistinguishable (see Figure~1). In turn, this suggests
that these two clusters represent an initial burst of star cluster
formation that extended over a huge volume of the proto-Galaxy. In view
of the fact (VandenBerg, Bolte and Stetson 1990) that all the most
metal-deficient halo clusters for which accurate relative ages are
available have the same age to within $\sim$0.5~Gyr, including now
NGC$\thinspace$2419 at nearly twice the distance of the Magellanic
Clouds in the far halo, we have a powerful constraint on the spatial
extent of the initial cluster formation epoch.

However, interpretation of preliminary HST data for Pal~4, another far
halo object which is thought to be a strong second-parameter cluster,
has proven problematical. Available metallicity determinations span
a wide range, which makes it difficult to select an appropriate well
studied inner halo cluster for differential comparison with Pal~4. If
we choose M5 for that purpose, initial results would suggest that Pal~4
is some 5~Gyrs younger via the $\Delta$color technique and about
1.5~Gyrs younger via the $\Delta$V technique. If the two clusters have
the same [Fe/H] but there were a dramatically different ratio of
[$\alpha$/Fe] in them, that might explain the discrepant relative ages
inferred by the two techniques.  We are also concerned that the Pal~4
reductions of data spanning two cycles may still have some small
photometric calibration error.

In summary, from the study of globular star clusters in the outer halo,
several important conclusions hold. New HST results suggest that the
initial burst of star cluster formation was a global, rapid phenomenon
that occurred over a huge volume. The absolute ages of these oldest,
metal-deficient clusters, 15$\pm$2~Gyr, conflict with most recent
determinations of H$_{\rm o}\sim70$. While at all R$_{\rm gc}$ and at
most metallicities there definitely are a handful of globular clusters
that appear younger by a few Gyr than the bulk of the system, much
recently accumulated evidence suggests that age as the dominant second
parameter remains open to debate.

\section {QUESTION 4}

It has been clear since the late 1970s and the work of Zinn, Searle
and Zinn, Aaronson and Mould, and others that the other bound
stellar systems in the outer halo, the dwarf spheroidal (dSph)
galaxies, contain more than one generation of stars. Their great
distances have made quantitative analysis slow going, but recent work
offers dramatic insight into the objects that some consider to be
possible building blocks for the Milky Way halo.

As shown in Figure 2, deep photometry from the CTIO 4-m telescope by
Smecker-Hane, et al.  (1994, 1997) for the Carina dSph reveals evidence
for four distinct episodes of star formation. Efforts by Smecker-Hane to
create synthetic CMDs for Carina suggest that some 2\% of the mass is
represented by star formation that ocurred 1 to 1.5 Gyr ago, 28\% of the
mass by star formation that ocurred from 2.5 to 3.5 Gyr ago, 50\% of the
mass by star formation about 3.5 to 7 Gyr ago, and the remaining 20\% of
the mass by star formation about 10 to 14 Gyr ago.  Amazingly, there is
very little evidence for chemical enrichment in excess of
$\sim$0.2-0.3~dex; it would appear that galactic winds must have let most
of the newly synthesized metals escape during the major star formation
epoch $\sim$6~Gyr ago while leaving a significant fraction of the gas
behind. The long hiatus between the initial and subsequent bursts of star
formation is puzzling; why wouldn't there have been star formation driven
by a recollapse on a cooling time scale of some 100 Myrs?

\begin{figure}[t]

\centerline{\epsfysize=8cm\epsfbox{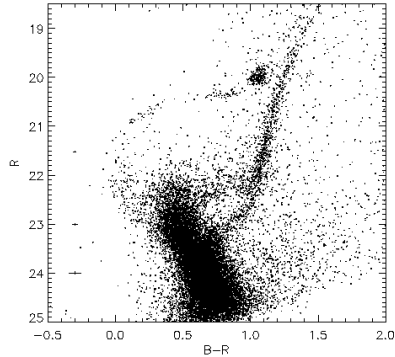}}
\caption{ A CMD for the Carina dwarf spheroidal galaxy
    (Smecker-Hane et al. 1997).} 
\end{figure}

Aaronson and Mould's (1980) discovery of carbon stars in the much more
luminous (M$_{\rm v}=-14$) dSph, Fornax, strongly hinted at an
intermediate age population. Fornax has five globular clusters,
which makes it unique among the classical outer-halo dSphs.  Buonanno,
et al. (1985) provided evidence for a large range in metallicities in
Fornax, $-2.2<{\rm [Fe/H]}<-0.7$.  CCD photometry from Dec. 1995 with the
CTIO 1.5-m telescope by Stetson, Hesser and Smecker-Hane (1997; see
Stetson 1996) covers a large portion of Fornax and reveals several
interesting properties.  Star formation appears to have been remarkably
steady until about 4~Gyr ago, when it dropped dramatically;
nonetheless, there is a smattering of stars only a few 100 Myrs old
(see also Beauchamp, et al. 1995)!  The red-giant and horizontal branches
show clearly the effects of metallicity evolution.  In
contradistinction to Carina, it is clear that Fornax has undergone a
complex evolution in age and metallicity.

Mighell \& Rich (1996) used HST to image the Leo~II dSph, from which
they infer a $\pm$0.25~dex range in [Fe/H] and a dominant stellar
population of age 9$\pm$1~Gyr,  although they note that the CMD
definitely shows a range of ages from 7 to 14 Gyr.  CFHT and KPNO 4-m
photometry of the Draco dSph (Stetson 1996) reveals yet another
distinct CMD, which this time appears to be dominated by a single old
population with a modest percentage of younger and/or peculiar stars.
Thus, the situation with dSphs is reminiscent of planetary satellites
in the solar system: just as Voyager images revealed each satellite to
be unique and the system to possess an unimagined rich spectrum of
characteristics, deep CMDs for the dSphs reveals them to be
unique and their system to be rich with clues to star formation
processes in low-mass galaxies.

\begin{figure}[t]
\centerline{\epsfysize=8cm\epsfbox{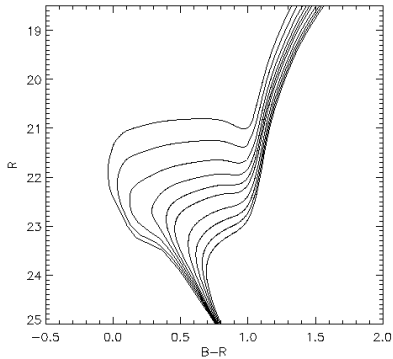}}
\caption{New theoretical isochrones for [Fe/H]$=-1.84$  from 
    (VandenBerg et al. 1997). For comparison with the Carina dSph CMD, 
    ages 1.3, 2, 3, 4, 5, 6, 8, 10, 12, 14 Gyr are shown.}
\end{figure}

When, as in the previous section, we focus attention upon the globular
star clusters in the Galactic halo, we develop a quite restricted sense
of the stellar populations there. The globulars, which represent only
some 2\% of the luminous mass in the halo, generally exhibit a modest
range, if any, in age from one cluster to another, and no detectable
range within a given cluster. On the other hand, dSphs, whose radial
distribution overlaps with the outermost globulars of the Galaxy,
exhibit episodic star formation, that, in the apparent absence of gas
(Knapp, Kerr \& Bowers 1978), is as if by magic.  [Note: the latter HI
observations definitely deserve repeating, because then poorly-known
radial velocities were used to determine the scanning range; the
resulting limits could be greatly improved.] Carina and Fornax offer
dramatially contrasting situations. Carina exhibits a large age spread
arising from distinct star formation episodes and a small [Fe/H] range,
while Fornax exhibits a large age spread, almost continous star
formation, and a large spread in [Fe/H].  Spectra are needed for large
samples of stars if we are to disentangle age and metallicity
degeneracies in the dSphs,  and  if we want to learn how the gas
content evolved with time (i.e., inflow vs outflow). The dSphs have
much to each us about how star formation is regulated, and how chemical
elements evolve in young, proto-Galactic fragments and also in low
metallicity quasar absorption line systems.

\section {QUESTION 5}

Age determinations for globular clusters, with their fascinating
implications for cosmology and the chronology of the formation of the
Galaxy, require knowledge of the abundances of the objects being
dated.  However, the uncertainties underlying our knowledge of halo
abundances are perhaps not stressed enough in our reach for answers to
the big questions.  Space does not allow treatment of this topic in
detail, and the reader is referred elsewhere for reviews (e.g., Briley
et al. 1994, Hesser 1996).  For instance, it is widely recognized that
in the halo, $\alpha$-elements are enhanced relative to the Fe-peak
elements, such that $ [\alpha/{\rm Fe}]\sim+0.4$ for [Fe/H]$<-0.5$.
Such trends presumably reflect the difference in the integrated
amount of chemicals produced from SNe I and II at different 
times. Similarly, the relative ratios of C, N \& O vary from cluster
to cluster and from star to star within individual clusters.  With present
technology, abundance determinations at spectral resolution above
40,000 (as required, for instance, to measure [O/Fe]) are limited to
the most luminous giants, whose C, N \& O behavior argues that they
have undergone deep mixing as they ascended the giant branch (see,
e.g., Briley et al. 1994). If the surface abundances have
been affected by dredge-up from the interior, what values should be
used for the age-sensitive turnoff from the main sequence when
comparing isochrones to observed CMDs?

In the past few years, measurements of the infrared triplet of Ca~II have
become synonomous with `metallicity' in many studies. Extensive
measurements of the Ca index for globular cluster giants (Rutledge et al.
1997a,b) found tight correlations with the widely used Zinn \& West (1984)
[Fe/H] scale based generally on integrated light and/or low-dispersion
spectroscopy, as well as with a scale based on sytematic analysis of
equivalent widths from high-dispersion spectra (Carreta \& Gratton 1996).
However, the latter authors demonstrate a significant non-linear
relationship beween their metallicity scale and that of Zinn \& West.
While there is no doubt that the Ca index is a powerful probe of halo
chemistry, it is perhaps not entirely clear to what extent it reveals
differences in $\alpha$ elements from cluster to cluster, differences in
[Fe/H] from cluster to cluster, or differences in Galactic formation
physics.

The challenges of making accurate measurements, and of then interpreting
them correctly through model atmospheres to deduce the chemistry of
stellar systems in the halo, leave open the nagging possibility that
perhaps we don't know their chemistry well enough to establish
definitive ages and an age profile for them. Conceivably some of the
answers to the fundamental questions of Galaxy formation and cosmology
addressed earlier are more compromised than we yet appreciate.
Moreover, to achieve a full understanding of the halo, we will need to
understand how the field and halo cluster stars came to have different
metallicity distribution functions (e.g., Suntzeff, et al. 1991, Laird,
et al., 1993, Carney, et al. 1996) and we will need accurate
abundances for the heavily reddened globulars in the innermost regions
of the Galaxy. Just as in making good kimch'i, we need to get the
chemistry right for the Galactic halo if we are to have confidence in
our derived properties...and we may have a greater distance to go towards
that goal than we'd like to believe.

\section {QUESTION 6}

Efforts to unravel the history of the formation of the Galaxy rely
heavily upon the stellar systems in the halo, yet, as noted earlier,
the globular clusters represent only some 2\% of the luminous mass
in the halo. A long standing question has been what fraction of the
halo field stars originated in globulars or other stellar systems
that might have dissolved due to Galactic tidal or other forces. 
A closely related question is what fraction of halo dark matter might 
be baryonic in the form of extremely low mass stars. As with the
previous section, the dynamical state and evolution of halo objects
is a vast subject, from which only a few salient points resulting
from recent studies can be made here.

Ground-based studies to faint limiting magnitudes for nearby globular
clusters raised the tantalizing possibility that the luminosity function
in some clusters continues to rise for M$_{\rm I}>8$ (e.g., Richer et al.
1991). However, HST photometry by Paresce et al.  1995 and Richer et al.
1995 reveals a turnover in the luminosity function, such that it now seems
less likely that present-day clusters harbor a wealth of potential dark
matter for contribution to the Galactic halo through evaporation.
Analyses of star counts in the Hubble Deep Field (e.g. M\'endez et al.
1996) find, as do the cited globular cluster studies, that the halo
luminosity function is smaller than what would be projected by
extrapolation of the disc luminosity function in the solar neighborhood.
Other indicators (e.g. Mould 1996; Alcock et al.\ 1997) also support a
turnover in the main-sequence luminosity function of the halo.

For a number of years, there has been considerable theoretical debate
regarding how much  of the original globular cluster population in the
Galaxy has been destroyed by dynamical friction, disc shocking and
evaporation. Recently there has been convergence on the view that a
substantial fraction of the original cluster population in our Galaxy
and in others has likely been dissipated by these processes, which act
with differing efficiencies depending upon the mass, concentration,
location and orbital parameters of individual clusters (e.g., Okazaki
\& Tosa 1995, Capriotti \& Hawley 1996, Gnedin \& Ostriker 1997).
Moreover, it is argued that over very long times these processes could
transform cluster systems with power-law luminosity functions and
young, massive clusters (like those found in the Magellanic Clouds)
into systems of old clusters with Gaussian luminosity functions and
other similar systemic properties as observed for many galaxies (e.g.
Harris 1991).

From the perspective of trying to understand the balance between
dissipative collapse of gas and accretion of previously formed stellar
systems in the formation of the Milky Way and other large galaxies, the
discovery of the dissolving Sagittarius dSph (Ibata, Gilmore \& Irwin
1994) provides reassurance (as if any were needed!) that mergers
occur.  On the other hand, Unavane, Wyse \& Gilmore (1996) use chemical
properties of the dSphs to place upper limits on the number of Carina
and Fornax dSphs that could have been accreted by the Milky Way in the
past 10~Gyrs. They argue, based upon the colour distribution and
metallicities of their stars compared with field halo stars, that fewer
than 60 Carina-like and fewer than six Fornax-like dSphs could have
been accreted in the past 10 Gyrs. However, potential complications in
an analysis like theirs may revolve around the balance between
Galactocentric distance (e.g., stripping of gas from a dSph) and
internal physical processes as regulators of star formation history.

Theoretical ideas, from Toomre \& Toomre (1972) to the present, about
the role of dynamical merging in galaxy formation are summarized by
Lynden-Bell \& Lynden-Bell (1995), whose new analysis of positions and
velocities for globulars and dSphs identifies numerous possible ghostly
streams in the halo that may mark orbits of ancestral satellites that
merged into the Milky Way. Johnston, Hernquist \& Bolte (1996) have
also demonstrated new analysis techniques that suggest `debris from
minor mergers can remain aligned along great circles throughout the
lifetime of the Galaxy,' while Majewski, Munn \& Hawley (1996) used 
a north Galactic pole proper motion survey to argue that the
halo is not dynamically mixed.  Bellazzini, et al. (1996) suggest that
inner halo GCs experienced a different, much more chaotic dynamical
evolution from that of disc GCs, even though the radial zone the two
groups occupy is nearly the same.  Minniti's (1996) kinematic study of
bulge red giants strengthens the conclusion that the Galactic bulge
formed via dissipative collapse from material remaining after halo
formation; among his predictions is that the metal-rich bulge clusters
should not be older than the metal-poor halo clusters.

In brief, then, the topic of `ghosts' in the Galactic halo touches upon
such fundamental issues as the role of baryonic material in dark matter
and the history Galactic formation. Recent discoveries of a turnover in
deep GC luminosity functions appears to diminish the likelihood that
evaporation of extremely faint stars from present GCs contributes in a
major way to halo dark matter. That notwithstanding, there has been
convergence recently in favor of the idea that the observed GCs are
likely a small fraction of the original population, such that a high
percentage of halo field stars might have originated in now-dissipated
GCs. Color and metallicity properties of field halo stars limit to
relatively few the number of dSphs that can have been accreted in the
last 10~Gyrs. New analytical techniques and models provide hope that
ethereal signatures of past merger events can be mapped, and thus 
that we may eventually be able to constrain the relative importance of
dissipative collapse and mergers in Galactic halo formation.

\section {SUMMARY}

Unavoidably a brief review that attempts to highlight major research
themes for a diverse audience  leaves out much and risks not conveying
the excitement and controversies of a subject as vast of the formation
of the Milky Way. Readers wishing to explore more will find thorough
reviews and additional references in, e.g., van der Kruit \& Gilmore
(1996), Blitz \& Teuben (1996), Morrison \& Sarajedini (1996) and van
den Bergh (1996).

To summarize the theme of this review, the collective evidence
presently favors the view that the Milky Way likely formed by a
combination of rapid, dissipative collapse (which may have been
particularly dominant in the inner regions) and mergers with stellar
systems such as dSphs and dIrrs (which may have been particularly
dominant in the outer regions). The balance between these processes is
a matter of intense current research and controversy. An important
constraint stemming from recent ground-based and HST-based studies is
that the GCs most deficient in heavy elements exhibit a range of ages
$<0.5$~Gyrs, whether located in the far outer halo or the inner halo.
That is, initial star and star cluster formation appear to have
occurred simultaneously over a huge volume. Such a remarkable
observation seems worthy of careful contemplation over \it kimch'i \rm
and \it soju \rm outside this beautiful lecture theatre.

\acknowledgements
J.E.H. wishes to thank Profs. Hyung Mok Lee and Hong Bae Ann for inviting
his participation, for financial support and, most of all, for having
organized a stimulating and exceptionally pleasant meeting for all
participants in the beautiful city of Pusan at the superb facilities of
PNU.

\end{document}